\documentclass[12pt]{article}

\ifx\pdfoutput\undefined
\usepackage[dvips,bookmarks]{hyperref}
\else
\usepackage{hyperref}
\fi
\hypersetup{colorlinks=false,bookmarksopen,bookmarksnumbered,citecolor=blue,
   pdfstartview=FitH}

\usepackage[dvips]{graphicx}
\usepackage{latexsym}

\usepackage{amssymb,amsfonts,amsmath}
\usepackage{graphicx} 
\usepackage{indentfirst}

 \usepackage{bbm}

\parskip=\medskipamount

\newcommand {\cL}{{\cal L}}
\newcommand {\cM}{{\cal M}}
\newcommand {\cN}{{\cal N}}

\newcommand {\cP}{{\cal P}}

\def\a{\alpha}

\def\d{\delta}
\def\e{\epsilon}
\def\f{\phi}

\def\G{\Gamma}

\def\o{\omega}

\def\q{\theta}

\def\s{\sigma}

\def\z{\zeta}

\def\F{\Phi}
\def\J{\Psi}

\def\O{\Omega}

\def\S{\Sigma}
\def\U{\Upsilon}
\def\X{\Xi}

\def\rd{{\rm d}}
\def\ri{{\rm i}}

\newcommand{\ve}{\varepsilon}                            

\newcommand{\pa}{\partial}                           
\newcommand{\hf}{\frac12}

\newcommand{\be}{\begin{equation}}
\newcommand{\ee}{\end{equation}}
\newcommand{\bea}{\begin{eqnarray}}
\newcommand{\eea}{\end{eqnarray}}

\def\dt#1{{\buildrel {\hbox{\LARGE .}} \over {#1}}}    

\newcommand{\bm}[1]{\mbox{\boldmath$#1$}}

\def\double #1{#1{\hbox{\kern-2pt $#1$}}}

\begin{document}
\begin{titlepage}
\begin{flushright}
October, 2011 \\
\end{flushright}
\vspace{5mm}

\begin{center}
{\Large \bf  Comments on \mbox{$\bm{ \cN=2}$}  supersymmetric \mbox{$\bm \s$}-models
in projective superspace}
\end{center}

\begin{center}

{\bf
Sergei M. Kuzenko\footnote{sergei.kuzenko@uwa.edu.au}
} \\
\vspace{5mm}

\footnotesize{
{\it School of Physics M013, The University of Western Australia\\
35 Stirling Highway, Crawley W.A. 6009, Australia}}  
~\\
\vspace{2mm}

\end{center}
\vspace{5mm}

\begin{abstract}
\baselineskip=14pt
For the most general off-shell  $\cN=2$ supersymmetric $\s$-model in projective superspace, 
we elaborate on its formulation in terms of $\cN=1 $ chiral superfields.
A universal (model-independent) expression is obtained for the holomorphic symplectic two-form,
which determines the second supersymmetry transformation. This two-form
is associated with the two complex structures of the hyperk\"ahler target space, which are complimentary 
to the one used to realize the target space as  a K\"ahler manifold. 
\end{abstract}
\vspace{1cm}

\vfill
\end{titlepage}

\renewcommand{\thefootnote}{\arabic{footnote}}
\setcounter{footnote}{0}




A few years ago, Lindstr\"om and Ro\v{c}ek \cite{LR2008}
uncovered fascinating 
properties of $\cN=2$ supersymmetric $\s$-models in projective superspace 
\cite{KLR,LR88,LR90},  building on the geometric methods developed some 
twenty years earlier in \cite{HitchinKLR}. 
Here we will show how to derive such properties 
by using only the considerations of supersymmetry and duality. 

The most general off-shell $\cN=2$ supersymmetric nonlinear $\s$-model 
in projective superspace \cite{LR88} can be realized in terms of polar supermultiplets
(see, e.g., \cite{K-lectures} for a review of the projective superspace approach).
The corresponding action can naturally be re-formulated in $\cN=1$ superspace and has the form \cite{LR88} 
\bea
S[\U, \breve{\U}]  =  
\frac{1}{2\pi {\rm i}} \, 
\oint \frac{{\rm d}\z}{\z} \,  
 \int 
 \rd^4 x\,{\rm d}^4\q
 \, 
\cL \big( \U^I  , \breve{\U}^{\bar{J}} , \z  \big) ~,
\label{nact} 
\eea
where $\z$ is the inhomogeneous complex coordinate for ${\mathbb C}P^1$,  
and the arctic $\U (\z)$ and  antarctic $\breve{\U} (\z)$ dynamical variables  
are generated by an infinite set of ordinary $\cN=1$ superfields:
\be
 \U (\z) = \sum_{n=0}^{\infty}  \, \U_n \z^n = 
\F + \S \,\z+ O(\z^2) ~,\quad
\breve{\U} (\z) = \sum_{n=0}^{\infty}  \, {\bar
\U}_n
 (-\z)^{-n}=\bar \F -\frac{1}{\z} \bar \S +O(\z^{-2}) ~.
\label{exp}
\ee
Here $\F:=\U_0 $ is chiral, $\bar D_{\dt \a} \F =0$, $\S:=\U_1 $  is complex linear, $\bar D^2 \S=0$, 
while the remaining components, $\U_2, \U_3, \dots, $ are unconstrained complex 
$\cN=1$ superfields.  
The latter superfields are auxiliary, since they  appear in the action without derivatives. 

Although the $\s$-model  \eqref{nact} was first
introduced in 1988  \cite{LR88},
for some ten years it remained  a purely formal construction, because  
there existed no technique to eliminate the auxiliary superfields contained in $\U^I$, 
except in the case of  Lagrangians quadratic in $\U^I $ and $\breve{\U}^{\bar I}$.
This situation changed  in the late 1990s when  refs. \cite{K98,GK1,GK2} 
identified a subclass of  models (\ref{nact}) with  interesting geometric properties.
They correspond to the special case
\bea
\cL \big( \U^I  , \breve{\U}^{\bar{J}} , \z  \big) = K \big( \U^I  , \breve{\U}^{\bar{J}}  \big) ~,
\label{nozeta}
\eea
where $K \big( \F^I  , \bar \F^{\bar{J}}  \big)$ is the K\"ahler potential of a K\"ahler manifold $\cP$.
The target space of the corresponding $\cN=2$ supersymmetric $\s$-model  \eqref{nact} can be shown to be 
(a neighborhood of the zero section of)  the cotangent bundle $T^*\cP$  \cite{K98,GK1,GK2}.
For these models one can develop a simple procedure to eliminate the auxiliary superfields in 
perturbation theory, and exactly in the case when $\cP$ is an arbitrary Hermitian symmetric space
\cite{GK1,GK2,AN,AKL1,AKL2,KN}. For the $\s$-model associated with 
an arbitrary $K \big( \F  , \bar \F  \big)$, 
one can also develop its formulation in terms of $\cN=1 $ chiral superfields by making use of the considerations 
of supersymmetry and duality \cite{K-hyper,K-duality}. Here we extend the approach of \cite{K-duality}
to the case of arbitrary $\cN=2$ supersymmetric $\s$-models  (\ref{nact}). 

We assume that the  the auxiliary superfields in the model  (\ref{nact}) have been eliminated. 
Then,
the action  (\ref{nact})  turns into
\bea
S= \int  \rd^4 x\,{\rm d}^4\q \, {\mathbb L} (\F^I,  \S^J , \bar \F^{\bar I} ,\bar \S^{\bar J} )~,
\label{on-shell-action}
 \eea
 for some Lagrangian ${\mathbb L}$.

The action (\ref{nact}) is manifestly $\cN=1$ supersymmetric, and is also
invariant under the off-shell  second supersymmetry 
transformation \cite{LR88} (see also \cite{K-hyper} for a detailed derivation): 
\begin{subequations}
\bea
\d \U_0 &=& {\bar \ve}_{\dt \a} {\bar D}^{\dt \a} \U_1 ~,
\qquad 
\d \U_1 =-\ve^\a D_\a \U_0 
+   {\bar \ve}_{\dt \a}{\bar D}^{\dt \a} \U_2 
~,  \label{arctic7} \\
\d \U_k &=&-\ve^\a D_\a \U_{k-1} 
+{\bar \ve}_{\dt \a} {\bar D}^{\dt \a} \U_{k+1} 
~, \qquad k>1~.
\label{arctic8} 
\eea
\end{subequations}
Upon elimination of the auxiliary superfields, 
this  supersymmetry transformation takes the form
\bea
\d \F^I &=& {\bar \ve}_{\dt \a} {\bar D}^{\dt \a} \S^I ~,
\qquad 
\d \S^I =-\ve^\a D_\a \F^I 
+   {\bar \ve}_{\dt \a}{\bar D}^{\dt \a} \U^I_2 \big(\F,  \S , \bar \F,\bar \S \big)
~,  
\label{arctic9} 
\eea
where $\U_2$ is now a composite field. 
In general, the explicit form of $ \U^I_2 \big(\F,  \S , \bar \F,\bar \S \big)$ is not known.
However, if eq. \eqref{nozeta} holds and $K \big( \F , \bar \F  \big)$ is the K\"ahler potential of an
Hermitian symmetric space, then it can be shown \cite{GK1,GK2} that 
\bea
\U^I_2 (\F,  \bar \F, \S , \bar \S)&=& -\hf \G^I_{JK} \big( \F, \bar{\F} \big) \, \S^J\S^K~,
\eea
with $\G^I_{JK} 
( \F , \bar{\F} )$  the Christoffel symbols for the  
K\"ahler metric $g_{I \bar J} ( \F , \bar{\F} )$.

It can be shown that the action (\ref{on-shell-action}) is invariant under the transformation
(\ref{arctic9}) if the following conditions hold:
\begin{subequations} \label{6}
\bea
\frac{\pa {\mathbb L}}{\pa \F^I}+ \frac{\pa {\mathbb L}}{\pa \S^J} \,\frac{\pa \U_2^J}{\pa { \S}^{I} } 
&=& \frac{ \pa \X}{ \pa { \S}^{ I} } ~,
\label{master2} \\
\frac{\pa {\mathbb L}}{\pa \S^J} \,\frac{\pa \U_2^J}{\pa {\bar \S}^{\bar I} } 
&=& \frac{ \pa \X}{ \pa {\bar \S}^{\bar I} } ~, \\
-\frac{\pa {\mathbb L}}{ \pa {\bar \S}^{\bar I} }
+ \frac{\pa {\mathbb L}}{\pa \S^J} \,\frac{\pa \U_2^J}{\pa {\bar \F}^{\bar I} } 
&=& \frac{\pa \X}{\pa {\bar \F}^{\bar I} }~,
\label{master3}
\eea
\end{subequations}
for some function $\X ( \F ,  \S, \bar \F, \bar \S)$ which is determined, up to a constant, 
in terms of $\mathbb L$ and $\U^I_2$.
In the case \eqref{nozeta}  the equations \eqref{6}  can be used to determine ${\mathbb L}$ and $\U^I_2$
 \cite{K-hyper,K-duality}. 

To uncover the explicit structure of the hyperk\"ahler target space associated with the $\s$-model
(\ref{on-shell-action}), we should construct a dual formulation of the theory (\ref{on-shell-action}) 
obtained by dualizing each complex linear superfield
$\S^I$ and its conjugate $\bar \S^{\bar I}$ into a chiral--antichiral pair $\J_I$ and $\bar \J_{\bar I}$.
This is accomplished through the use of the first-order action
\bea
S_{\text{first-order}}=   \int \rd^4 x\,{\rm d}^4\q \, 
\Big\{ {\mathbb L}\big(\F, \bar \F, \S , \bar \S \big)
+\J_I \,\S^I + {\bar \J}_{\bar I} {\bar \S}^{\bar I} 
\Big\}~.
\label{f-o}
\eea
Here  $\S^I$ is  an unconstrained complex superfield, 
while  $\Psi_I$ is chiral, 
${\bar D}_{\dt \a} \J_I =0$.
This model is equivalent to  (\ref{on-shell-action}). Indeed, varying $S_{\text{first-order}}$
with respect to $\J^I$  gives $\bar D^2 \S^I =0$ 
and then (\ref{f-o}) reduces to the original theory, eq (\ref{on-shell-action}). 
On the other hand, we can integrate out $\S$'s and their conjugates using their equations of motion
\bea
\frac{\pa  }{\pa \S^I}  {\mathbb L}\big(\F, \bar \F, \S , \bar \S \big)+ \J_I =0~,
\eea 
which can be used to express $\S$'s and their conjugates in terms of the other fields, 
$\S^I = \S^I (\F, \J, \bar \F, \bar \J)$.
This leads to the dual action
\bea
S_{\text{dual}}  
&=&   \int \rd^4 x\,{\rm d}^4\q \, {\mathbb K} \big(\F^I,  \J_J , \bar \F^{\bar I} ,\bar \J_{\bar J} \big)~.
\label{act-ctb}
\eea
It can be shown that the second supersymmetry of the original theory, eq.  \eqref{arctic9}, 
turns into the following  symmetry   of the first-order action \eqref{f-o}: 
\begin{subequations}
\bea
\d \F^I &=&\phantom{-}\hf {\bar D}^2 \Big\{ \bar{\e} \bar{  \q} \, \S^I \Big\} ~,  \\
\d \S^I &=&-\ve^\a D_\a \F^I 
+   {\bar \ve}_{\dt \a}{\bar D}^{\dt \a} \U^I_2 \big(\F,  \S , \bar \F,\bar \S \big)~,\\
\d \J_I &=&- \hf {\bar D}^2 \Big\{ \bar{\e}\bar{ \q} \, \frac{\pa \mathbb L}{\pa \F^I} \Big\} ~  .
\eea
\end{subequations}
Using the standard properties of the Legendre transformation, 
we then derive the second supersymmetry transformation of the dual theory \eqref{act-ctb}. 
It has the form
\bea
\d \F^I &=&\hf {\bar D}^2 \Big\{ \bar{\e} \bar{\q} \, \frac{\pa \mathbb K}{\pa \J_I} \Big\} ~, 
\qquad
\d \J_I =- \hf {\bar D}^2 \Big\{ \bar{\e} \bar{\q} \, 
\frac{\pa \mathbb K}{\pa \F^I}   \Big\}~.
\label{SUSY-ctb4}
\eea
Finally, if we introduce the condensed notation 
$\f^a := (\F^I\,, \J_I)$ and  ${\bar \f}^{\,\bar a} = ({\bar \F}^{\bar I}\,, {\bar \J}_{\bar I})$, as well as
the symplectic matrices
\bea
{\mathbb J}^{a b} = {\mathbb J}^{\bar a \bar b} = 
\left(
\begin{array}{rc}
0 ~ &  {\mathbbm 1} \\
-{\mathbbm 1} ~ & 0  
\end{array}
\right)~,  
\qquad 
{\mathbb J}_{a b} = {\mathbb J}_{\bar a \bar b} = 
\left( 
\begin{array}{rc}
0 ~ &  {\mathbbm 1} \\
-{\mathbbm 1} ~ & 0  
\end{array}
\right)~, 
\eea
then the supersymmetry transformation (\ref{SUSY-ctb4}) can be rewritten as 
\bea 
\d \f^a &=&\hf 
{\bar D}^2 \Big\{ \bar{\e} \bar{\q} \,  {\mathbb J}^{ab}  \frac{\pa \mathbb K}{\pa \f^b} \Big\} ~.
\label{SUSY-ctb5}
\eea

At this stage we should recall the classic results of \cite{HKLR} on the formulation of the most general 
$\cN=2$ supersymmetric $\s$-model in terms of $\cN=1$ chiral superfields. Let us denote by $\cM$ 
the hyperk\"ahler target space of such a $\s$-model. To realize the $\s$-model in $\cN=1$ superspace, 
we have to pick out an integrable  complex structure on $\cM$, say $J_3$, and introduce associated local complex coordinates $\f^a$ in which $J_3$ takes the form
\begin{align}\label{complex_structure1} 
J_3 = \left(\begin{array}{cc}
\ri \,\delta^a{}_b & 0 \\
0 & -\ri \,\delta^{\bar a}{}_{\bar b}
\end{array}\right)~.
\end{align}
The other complex structures, $J_1$ and $J_2$, can be chosen as 
\begin{align}
\label{complex_structure2}
J_1 = \left(\begin{array}{cc}
0 & g^{a \bar c} \bar \o_{\bar c \bar b} \\
g^{\bar a c} \omega_{cb} & 0
\end{array}\right), \qquad
J_2 = \left(\begin{array}{cc}
0 & \ri\, g^{a \bar c} \bar \o_{\bar c \bar b}   \\
-\ri\, g^{\bar a c} \omega_{cb} & 0
\end{array}\right)~,
\end{align}
where $\o_{ab} (\f) $ and $\bar \o_{\bar a \bar b} (\bar \f )$ are covariantly constant two-forms  
on $\cM$ such that 
\begin{align}
\omega^{ac} \omega_{cb} = -\delta^a{}_b~, \qquad
\o^{ab} (\f) := g^{a\bar c} (\f, \bar \f) g^{b \bar d} (\f, \bar \f) \, \bar \o_{\bar c \bar d}(\bar \f)~.
\label{3.18}
\end{align}
We denote by ${\mathbb K}\big(\f^a, {\bar \f}^{\overline{b}}\big)$ 
the K\"ahler potential on $\cM$ with respect to $J_3$.
Then the $\cN=2$ supersymmetric $\s$-model is described by the action 
\bea
S&=& \int {\rm d}^4 x \,{\rm d}^4 \q  \, {\mathbb K}\big(\f^a, {\bar \f}^{\overline{b}}\big)~,
\qquad {\bar D}_{\dt \a} \f^a =0~. 
\label{N=1sigma-model}
\eea
It is invariant under a  second supersymmetry which, modulo a trivial symmetry transformation, 
can be chosen in the form \cite{HKLR}
\bea
\d \f^a &=&\hf 
{\bar D}^2 \Big\{ \bar{\e} \bar{\q} \,\o^{ab}    \frac{\pa \mathbb K}{\pa \f^b} \Big\}~.
\label{barOmega}
\eea

Comparing the relations \eqref{SUSY-ctb5} and \eqref{barOmega} gives
\bea
\o^{ab} (\f) = {\mathbb J}^{ab}\quad \Longrightarrow \quad \o_{ab} (\f) = {\mathbb J}_{ab}~.
\eea
We conclude that the holomorphic symplectic two-form
${\o}^{(2,0)}$ of the hyperk\"ahler target space coincides with 
the canonical holomorphic symplectic two-form, 
\bea
{ \o}^{(2,0)} := \hf { \o}_{ab} \,{\rm d}\f^a \wedge {\rm d} \f^b
 =  {\rm d} \F^I  \wedge{\rm d}\J_I~.
 \label{omega-j}
\eea
The chiral superfields $\F^I$ and $\J_I$ are Darboux coordinates for ${ \o}^{(2,0)} $.

The above consideration shows how to reconstruct, in principle,  
an $\cN=2$  supersymmetric $\s$-model 
of the type  (\ref{on-shell-action}) starting from an arbitrary hyperk\"ahler manifold $\cM$. 
This requires implementing the following steps: (i) pick out a complex structure $J_3$ 
to realize $\cM$ as the K\"ahler manifold with respect to $J_3$; (ii) introduce Darboux coordinates
for ${ \o}^{(2,0)}$ and then construct the $\s$-model (\ref{act-ctb}); (iii) perform the inverse Legendre
transformation (or the chiral-linear duality, following the terminology of \cite{KLvU})
with respect to the variables $\J_I$ and their conjugates. However, our consideration 
seems to give no clue as to how to reconstruct the off-shell action (\ref{nact}) except for the case when 
the Lagrangian in  (\ref{nact}) has no explicit $\z$-dependence, eq. \eqref{nozeta}. 
In that case, the hyperk\"ahler potential has the general form (see, e.g., \cite{K-duality})
\bea
{\mathbb K} \big(\F,  \J , \bar \F ,\bar \J \big) &=& K \big( \F, \bar{\F} \big)
+ \sum_{n=1}^{\infty} H^{I_1 \cdots I_n {\bar J}_1 \cdots {\bar 
J}_n }  \big( \F, \bar{\F} \big) \J_{I_1} \dots \J_{I_n} 
{\bar \J}_{ {\bar J}_1 } \dots {\bar \J}_{ {\bar J}_n } ~,
\eea
where $H^{I {\bar J}} \big( \F, \bar{\F} \big) 
= g^{I {\bar J}} \big( \F, \bar{\F} \big) $ and 
 the coefficients $H^{I_1 \cdots I_n {\bar J}_1 \cdots {\bar J}_n }$, for  $n>1$, 
are tensor functions of the K\"ahler metric
$g_{I \bar{J}} \big( \F, \bar{\F}  \big) 
= \pa_I 
\pa_ {\bar J}K ( \F , \bar{\F} )$,  the Riemann curvature $R_{I {\bar 
J} K {\bar L}} \big( \F, \bar{\F} \big) $ and its covariant 
derivatives. 

A remarkable result of 
Lindstr\"om and Ro\v{c}ek \cite{LR2008} is that they showed that 
the $\cN=2$ superfield Lagrangian in  (\ref{nact}) can be identified 
with the generating function of a twisted symplectomorphism 
associated with the symplectic two-form
\bea
\O = { \o}^{(2,0)}  + \z { \o}^{(1,1)} -\z^2 { \o}^{(0,2)} ~,
\eea
where $ { \o}^{(1,1)}$ is the K\"ahler form associated with $J_3$ and 
${ \o}^{(0,2)} := \hf \bar{ \o}_{\bar a \bar b} \,{\rm d} \bar \f^{\bar a} \wedge {\rm d} \bar \f^{\bar b}$. 
This allows in principle to restore the Lagrangian in  (\ref{nact}) starting from $\cM$, 
however  a constructive procedure is not yet available.

The analysis of this paper can naturally be extended to the case of massive $\cN=2$ supersymmetric $\s$-models 
in projective superspace \cite{K-massive}.
\\

\noindent
{\bf Acknowledgements:}\\
The author is grateful to Daniel Butter and Joseph Novak for reading the manuscript.
This work is supported in part by the Australian Research Council.

\small{

}

\end{document}